\documentstyle[nato]{crckapb}
\begin{opening}
\title{ON OBSERVING THE COSMIC MICROWAVE BACKGROUND}

\author{L.A. PAGE}
\institute{Princeton University\\
           Dept. of Physics\\
           Princeton, NJ}

\end{opening}

\runningtitle{COSMIC MICROWAVE BACKGROUND}

\begin{document}
%
%
\input epsf


\begin{abstract}

	The cosmic microwave background (CMB) comprises the
oldest photons in the universe and is arguably our most direct
cosmological observable. All precise and accurate measurements of
its attributes serve to distinguish between cosmological models.
Detector technology and observing techniques have advanced to the
point where fluctuations in the CMB of order a few microkelvin
are measured almost routinely. In these lecture notes, we review
recent measurements of both the absolute temperature and the
anisotropy of the CMB and discuss the relation between the data
and the general theoretical framework. Future directions are
indicated and the upcoming satellite experiments are discussed.

\end{abstract}

\section{Introduction}

	The CMB is a powerful probe of cosmology because 
essentially no steps separate what is measured from what is of
cosmological import; what you see is what you get. The CMB
photons have free-streamed through the cosmos since last
scattering off electrons some 100,000 years after the big bang.
The spectrum of the CMB is indistinguishable from a Planck
function to roughly 0.01\% accuracy. This tells us that there
were not any  highly energetic cosmic processes, that coupled to
photons,  before $z\approx10^3$. The near perfect shape of the
spectrum is the strongest evidence to date that the hot big-bang
model is correct. The pattern of minute spatial variations or anisotropy in
the CMB, which are of order $\delta T/T\approx 10^{-5}$, is a
fossil of the early universe. Furthermore, most models that give rise to
cosmic structure affect the CMB, leaving an imprint of a small
temperature fluctuation. 

	There are many review articles on both the spectrum of
and anisotropy in the CMB. For the anisotropy see [1]-[6];  for a
recent review of both the spectrum and anisotropy results see
\cite{SS96} and \cite{Page96}; and for reviews of the theory and
results on the spectrum see \cite{SZ80}, \cite{Danese90}, or
\cite{Bartlett91}. In addition, Partridge has written a new
book \cite{Partridge95} devoted to the subject.

%
%

	The outline for these notes is as follows. We discuss the 
microwave/far-infrared sky in Section 2. Next, in Section 3, we
discuss the recent results of absolute temperature  measurements
of the  CMB. After a model of the anisotropy is developed in
Section 4, Section 5 provides a discussion of the measurement
technologies and an overview of the canonical formalism for
describing the anisotropy. In Section 6, we review the state of
the field. We end with a discussion of the upcoming satellite
missions in Section 7.

\section{The Microwave/Far-Infrared Sky}

	The CMB is the brightest broad-band diffuse emitter in the sky
between about 1 and 500 GHz,  completely dominating the Galactic 
foreground emission. At the low frequency end of this range,
Galactic synchrotron emission exceeds the CMB and at the high
frequency end, interstellar dust emission dominates. Galactic
bremsstrahlung, or free-free emission, is the largest foreground
near 90 GHz. Figure 1 shows the frequency spectrum of the 
diffuse Galactic emission between 3 and
3000 GHz near a Galactic latitude of $20^{\circ}$. There are two 
sources not shown on this plot. Near 3000 GHz, thermal
emission from the interplanetary dust (Zodiacal light) is roughly
ten times smaller than the interstellar dust and its brightness scales with
frequency as $\nu^4$. Throughout the plotted
range, interstellar molecular line emission is also observed 
\cite{Wright91}. 

\begin{figure}
\epsfysize=3.5truein
\epsfverbosetrue
\centerline{\epsfbox{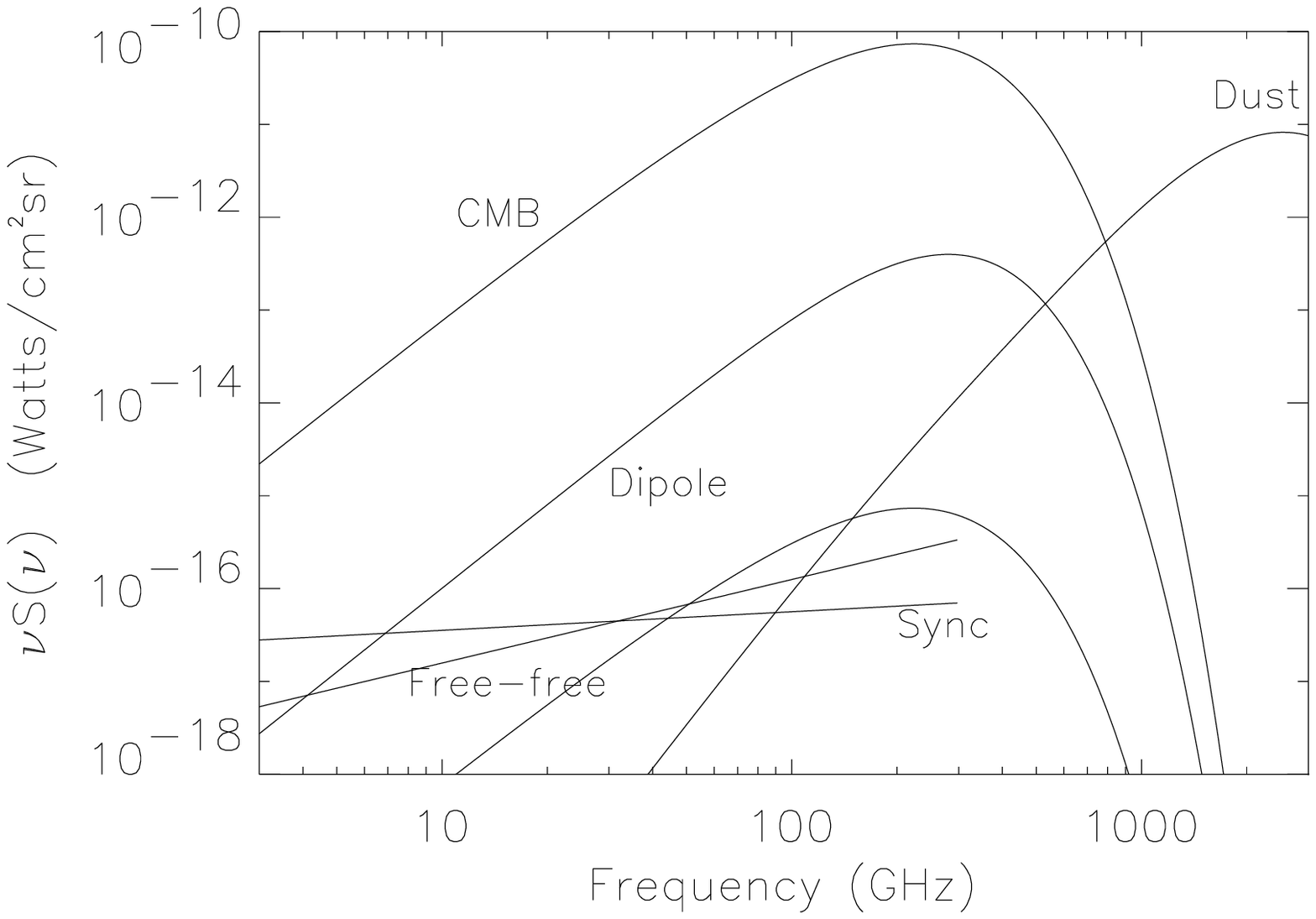}}
\begin{description}
\item[Figure 1.] The microwave sky from 3 to 3000 GHz near a
Galactic latitude of $b=20^{\circ}$. The ordinate
is the brightness of the sky times the frequency. With this 
convention, the plot indicates the distribution of power. For synchrotron
emission $S(\nu)\propto \nu^{-0.7}$; for free-free emission $S(\nu)\propto
\nu^{-0.1}$; and for dust emission near 100 GHz $S(\nu)\propto
\nu^{3.7}$. The scaling in effective temperature is 
$T(\nu )\propto\nu^{-2}S(\nu )$. The lowest Planck-like curve is
for ${\rm T} = 27~\mu$K.
\end{description}
\end{figure}

	In addition to the diffuse foreground,  galactic and
extragalactic point-like objects such as quasars, blazars,
gigahertz-peaked sources, and loud radio sources also emit
microwave and far-infrared radiation. The spectrum of  most
sources falls with increasing frequency, though not of every
source.\footnote{For sources, ``flat spectrum'' means that
S($\nu$) is independent of $\nu$, similar to free-free emission.} 
Relatively little is know about sources near 90 GHz. There are no
deep unbiased surveys from which to ascertain the ensemble
properties with confidence. Also, we know  that many of the
extragalactic high-frequency sources are variable.

	In Figure 1 it is evident that to probe either the spectrum
or the anisotropy to a part in 10$^{3}$ or 10$^{5}$ the
foreground emission must be confronted. Before COBE\cite{COBE},
the best full-sky maps were the ``Haslam {\it et al.}'' map at
408 MHz\cite{Haslam82} and the IRAS dust map, with Zodiacal light
subtracted, at 3000 GHz ($100~\mu$m)\cite{IPAC97}. Extrapolation
of these maps to frequencies of interest is problematic because
the spectral index varies from place to place on the sky and
there are components, such as free-free and cold dust emission,
that are missing from these maps. A believable CMB spectrum or
anisotropy experiment {\it must} measure the CMB and foreground
emission to similar precision.

	In addition to the frequency dependence of the foreground
emission, there is also a spatial dependence. At the largest 
angular scales, Galactic emission  falls off with Galactic
latitude roughly as $1/\sin(b)$. In other words, its intensity 
distribution, to first order, resembles a quadrupole: hot around
a circle, cold near the poles. Indeed this quadrupole confounded
some of the early measurements of the anisotropy and  its removal
from the {\sl COBE/DMR} data requires ingenuity
\cite{Kogut96}.

	The spatial distribution of celestial sources is 
commonly quantified with the angular spectrum. This formalism is
used to describe the anisotropy, radio sources, and the diffuse
foreground. It allows a direct comparison of the contribution from
each. To illustrate it, the relatively simple case of foreground
emission by extra-galactic radio sources is considered. We assume
radio sources are randomly placed on the sky and so the angular
spectrum should be that of ``white noise.'' If we measure the
total source emission temperature in each sky pixel, then that
distribution may be represented as an expansion in spherical
harmonics,

\begin{equation} 
T(\hat x ) = \sum_{l,m}a_{lm}Y_{lm}(\hat x ),
\end{equation}
where $\hat{x}$ is the unit vector in some direction. At this
time we will ignore the effects of finite measurement
resolution; this will be considered later. The
correlation function (or two-point function) of the temperatures is defined as
 
\begin{equation} 
C(\theta ) = <T(\hat{x} )T(\hat{y} )>,  ~~~{\rm
where}~~~\theta=\hat{x}\cdot\hat{y},
\end{equation}
and where the diagonal bracket indicates an ensemble average over many
universes. For small angular scales, the ensemble average can be replaced by an
average over positions in our universe. For a Gaussian random
field,  all the
information is contained in the two-point correlation function
(eq.\ 2). We do not know if the CMB is a Gaussian field, and we 
know that Galactic emission is certainly not. However, we still
use the two-point function; a more complete description
would use the higher-point correlation functions. Because $C(\theta)$
depends only on the  angular separation between two directions,
it may be expanded in Legendre polynomials as  

\begin{equation} 
C(\theta )=\sum_{l}{2l+1\over 4\pi}C_lP_l(\cos(\theta))
~~~{\rm where}~~~C_l=<| a_{lm}|^2>.
\end{equation}

	Here the brackets denote averaging over the $2l+1$ values
of $m$. The variance of the pixel temperatures is given by
$C(0)$. Because $P_l(0) = 1$, the variance in each ``mode'' $l$
is just $(2l+1)C_l/4\pi$. For ``white noise,'' $C_l = const.$ We
may see this by considering that the magnitude squared of the
Fourier transform of a uniform distribution of sources will also
be a uniform distribution in 2-D Fourier space; there is no
preferred scale. Call the uniform value $U$. For small angles,
the conjugate variable to angular separation is $l+1/2$, where 
$l+1/2 \approx l = 1/\delta\theta$ where $\delta\theta$ is the
angular scale. Now, the variance for a band of modes of width
$\delta l$ at a radius $l$ is given by the area in Fourier space
or $U2\pi l\delta l$. In other words, the  $C_l \propto U =
const$.

	To visualize the angular spectrum (``power spectrum'' is
usually reserved for the $C_l$), one generally plots
$(l(2l+1)C_l/4\pi)^{1/2} = \delta T_l$ versus $l$.  The extra
factor of $l$ means that $\delta T_l$  is the root mean squared
fluctuation per logarithmic interval, $\delta l/l$. The results
from such an analysis for radio sources and Galactic  emission at
40 GHz are shown in Figure 2.

	From Figures 1 and 2, it is clear that the foreground
emission  must be well understood before the absolute CMB
spectrum is known to 0.01\% near 3 GHz or before the angular
spectrum is known to the  few percent level. One mitigating
factor for the anisotropy is that the foreground fluctuations add
to the signal in quadrature in the approximation that they are
random fields. The absolute  measurements do not enjoy
this benefit. We are a long way from understanding the Galactic
emission and radio sources. For a thorough up-to-date assessment,
see Tegmark and Efstathiou \cite{TE96}.

\begin{figure}
\epsfysize=3.truein
\epsfverbosetrue
\centerline{\epsfbox{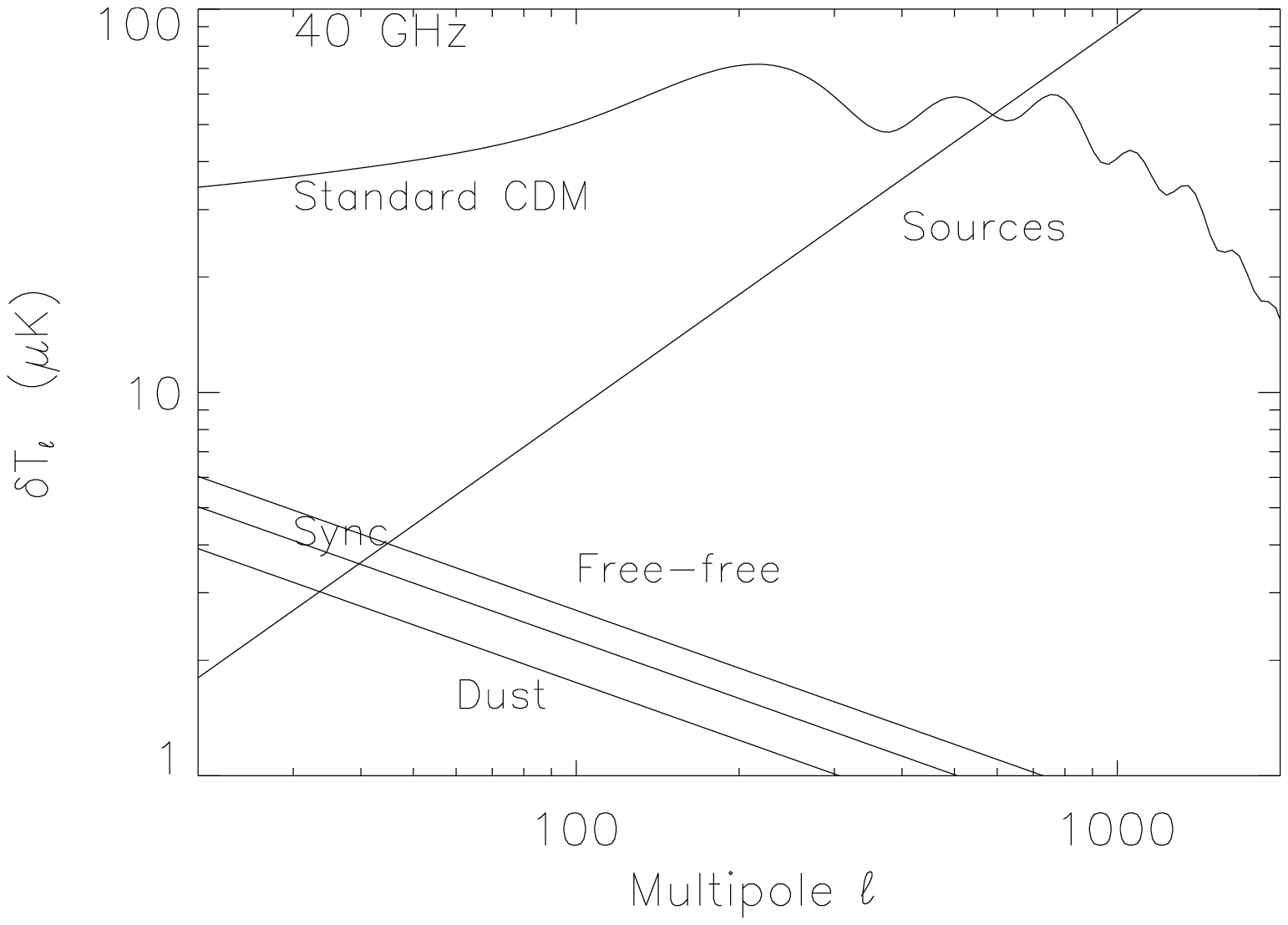}}
\begin{description}
\item[Figure 2.] Spatial spectrum of the foregrounds from
Netterfield {\it et al.} \cite{Netterfield97} and references
therein. This is for the
region  near the North Celestial Pole at a frequency of 40 GHz. 
At higher frequencies, the flux from dust increases and that from
synchrotron and  radio sources decreases. The curved line is
for standard CDM (from \cite{Rat95b}). 
\end{description}
\end{figure}

\section{Measurements of the Spectrum}

	The spectrum of the CMB is as close to that of a
blackbody as can be measured; no distortions have been detected. 
This is our best evidence that the universe went through a hot
dense phase when everything that interacts with photons was in
thermal equilibrium. In current models, the epoch of photon
production ended at $z\approx3\times10^{6}$, when the universe
was roughly two months old. In the subsequent expansion, an
injection of energy mediated by baryons would distort the
spectrum, though the injection would have to be large  (or
efficient) to be detectable today because there are roughly
$10^9$ photons per baryon. Thus a study of the spectrum is a study
of the history of cosmic energetics. 

	The FIRAS experiment \cite{Mather94} aboard the  {\sl
COBE} \cite{COBE} satellite measured the flux from the sky
between 2 and 96 cm$^{-1}$ ($60 - 2880$ GHz). Fixsen and
colleagues give the most recent results \cite{Fixsen96} and
discuss the exhaustive program  of systematic checks and
instrument calibration\cite{Fixsen94}. The FIRAS team finds that
the flux is described by the Planck function with a temperature
of

\begin{equation} 
{\rm T_{CMB} = 2.728\pm 0.004~K}.
\label{eq:1}
\end{equation}

One doubts that short of another satellite-based experiment  this
result will be matched or bettered at frequencies above 100 GHz.
It is comforting that the UBC rocket experiment \cite{Gush90}
gives a consistent result.

	The error on the FIRAS result, 4 mK, is
due entirely to systematic effects and Fixsen {\it et al.} interpret it as
a 95\% confidence limit. (The statistical error is $7~\mu$K.)
The error exists to tell the readers how confident the
authors are in the results. It should not be interpreted 
in the sense that ``if one hundred experiments were performed
only ten would lie outside the error bounds.''     

	When interpreting this result, one must bear in mind
that it comes from a {\it model} of the data. If the model is
not correct, the results must be re-interpreted. To be more
specific, the analyzed data come from maps of the sky in multiple frequency
bands between 2 and 21 cm$^{-1}$. A four parameter fit is made to
the maps at $|b|>5^{\circ}$ at each frequency,  

\begin{equation} 
{\rm FIRAS(\nu) = \alpha_0(\nu) Uniform + \alpha_1(\nu) Dipole 
+ \alpha_2(\nu) D9
+\alpha_3(\nu) D10}.
\label{eq:2}
\end{equation}

	The fit parameters scale the spatial distributions of a uniform
background, the CMB dipole, channel 9 from the DIRBE experiment
(D9, 72 cm$^{-1}$) and channel 10 from the DIRBE experiment 
(D10, 42 cm$^{-1}$). The last two maps are found to be good
measures of the  interstellar dust distribution. Two maps are
needed because there may be multiple dust components or, in an
explanation that Fixsen \cite{Fixsen96b} prefers, the dust temperature may be a
function of position. 

	The set of coefficients of the uniform component,
$\alpha_0(\nu)$, is then fit to a combination of four frequency
distributions. They are a) a blackbody, b) the derivative of a
blackbody (to fit an error to the temperature scale), c) the
spectrum of the Galaxy (that accounts for residual Galactic signal
in the monopole!), and d) a spectral distortion. The distortion
is parameterized by either a chemical potential $\mu$, or a
Compton $y-$factor. Only one distortion is fit at a time because
the spectral signatures of the two are anti-correlated. The
residuals to this fit and some of the basis functions are shown
in Figure 3.

\begin{figure}
\epsfysize=3.6truein
\epsfverbosetrue
\centerline{\epsfbox{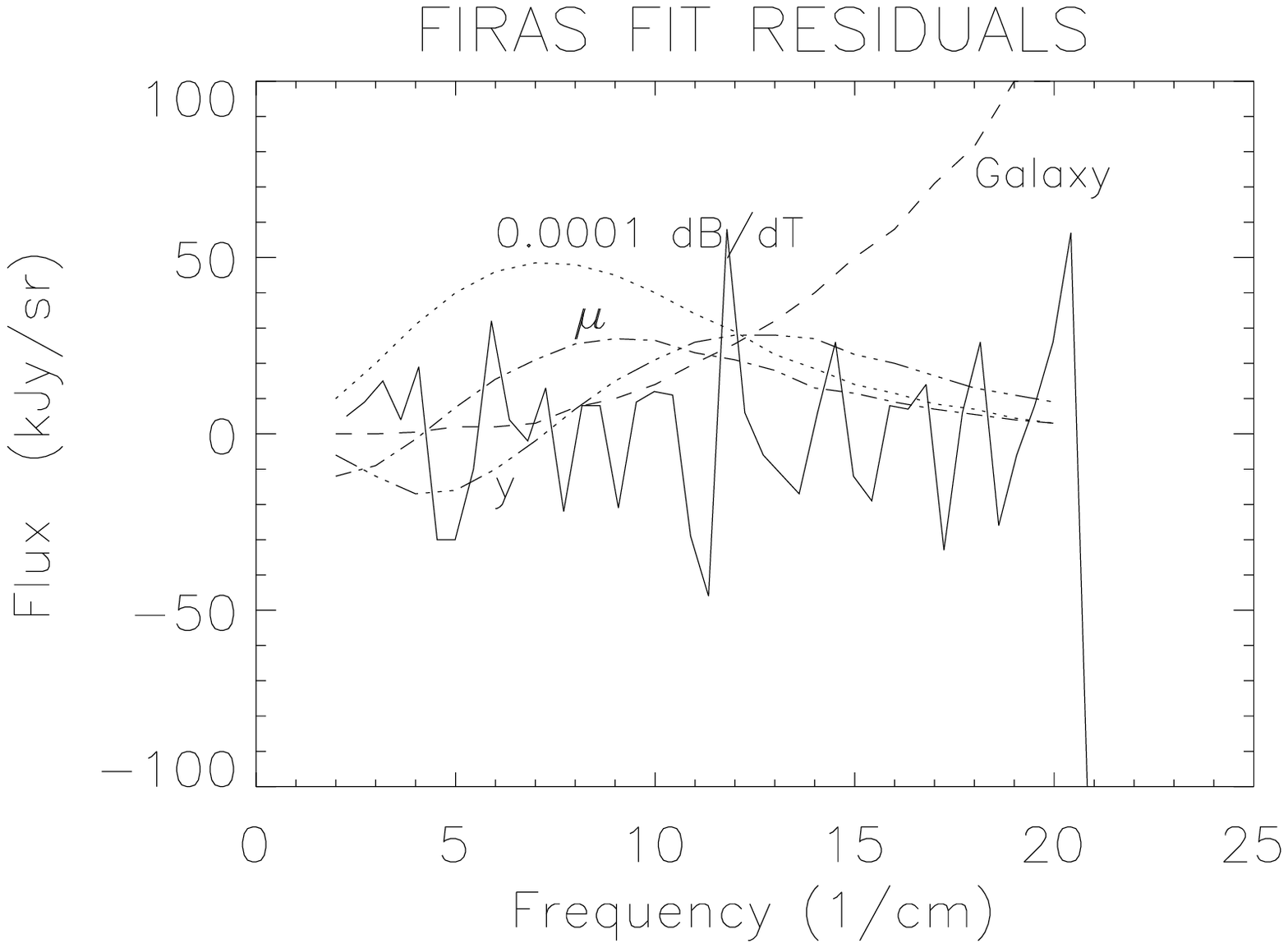}}
\begin{description}
\item[Figure 3.] Residuals of the fit to $\alpha_0(\nu)$ in 
eq.\ 5  from
\cite{Fixsen96}. If one starts with the $\alpha_0(\nu)$ in eq.\ 5
and subtracts a blackbody (curve not shown), a small calibration 
correction term (0.0001 dB/dT, where B$(\nu )$ is the Planck function), and a model of the Galactic
spectrum,  the solid line is obtained. The Galactic spectrum is
shown scaled to 1/4 its value at the galactic poles. In other
words, if a Galactic-type spectrum were not subtracted, the 
spectrum of the uniform sky map
would rise with frequency.  The peak of
the CMB is near $\nu = 5.5~\rm{cm}^{-1} = 165~$GHz. The intensity
measured there is roughly 385 MJy/sr. The $\mu$ and $y$ distortions are
shown at the 95\% CL values in eq.\ 6. The units on the
abscissa are converted to GHz by multiplying by 30. 
\end{description}
\end{figure}

	Before $z\approx3\times 10^{6}$,
double Compton scattering  and free-free emission 
maintain the thermal equilibrium of the CMB with the surroundings
by creating photons. An energy input simply results in a
hotter CMB temperature. Between $10^{5} < z < 3 \times
10^{6}$,  single Compton scattering, which conserves the number
of photons, is the dominant scattering mechanism over most of the
frequency spectrum. In this epoch, the CMB is in {\it
statistical} equilibrium  with its surroundings and the distribution is
characterized by a chemical potential $\mu$. (The quoted numbers
are for the unitless chemical potential; the flux is
$S_\nu(T,\mu)=2h\nu^3/[\exp(h\nu/kT_{CMB} +\mu)-1]$). At long
wavelengths, free-free emission is still effective and ``fills in
the tail'' of the distribution. For $z<10^{5}$, hot electrons,
which are neither in statistical nor in thermal equilibrium with the
CMB, can inverse Compton scatter the CMB photons to produce a
Compton $y$ distortion. When there are relatively few scattering
events, one may think of $y$ as the average fractional energy
change per scattering event times the average number of
scatterings \cite{RL79}, or $y=1/m_e c^2\int [k(T_e-T_{CMB})]
d\tau_e$ \cite{SZ80} where $T_e$ is the electron temperature and
$\tau_e$ is the optical depth due to scattering. Finally, if the
universe is ionized at $z<10^{3}$ then there may be enough
free-free emission from the plasma to increase the photon
occupation number at large $\lambda$ and thus the temperature
there. This distortion is parameterized by  $Y_{ff} = (h\nu
/kT)^2[T_{eff}(\nu )-T_{CMB}]/T_{CMB}$, with $T_{eff}$ the plasma
temperature.  These and other distortions, along with their
interpretation, are discussed in \cite{SZ80}, \cite{Bartlett91},
\cite{Wright94}, \cite{Bond95}.

	The best limits on $y$ and $\mu$ come from FIRAS
\cite{Fixsen96}. From these, Wright {\it et al.} \cite{Wright94}
constrain energy injection in the early universe as shown in 
Figure 4. The limit on $Y_{ff}$ \cite{Bersanelli94}
comes from a fit of the low frequency data.  The limits are:

\begin{eqnarray}
|y|      &<& 1.5\times10^{-5}~~95\%~~ CL,  \nonumber\\
|\mu|    &<& 9\times10^{-5}~~95\%~~ CL, \nonumber\\
Y_{ff}   &<& 1.9\times10^{-5}~~95\%~~ CL.
\label{eq:3}
\end{eqnarray}           

\begin{figure}
\epsfysize=3.5truein
\epsfverbosetrue
\centerline{\epsfbox{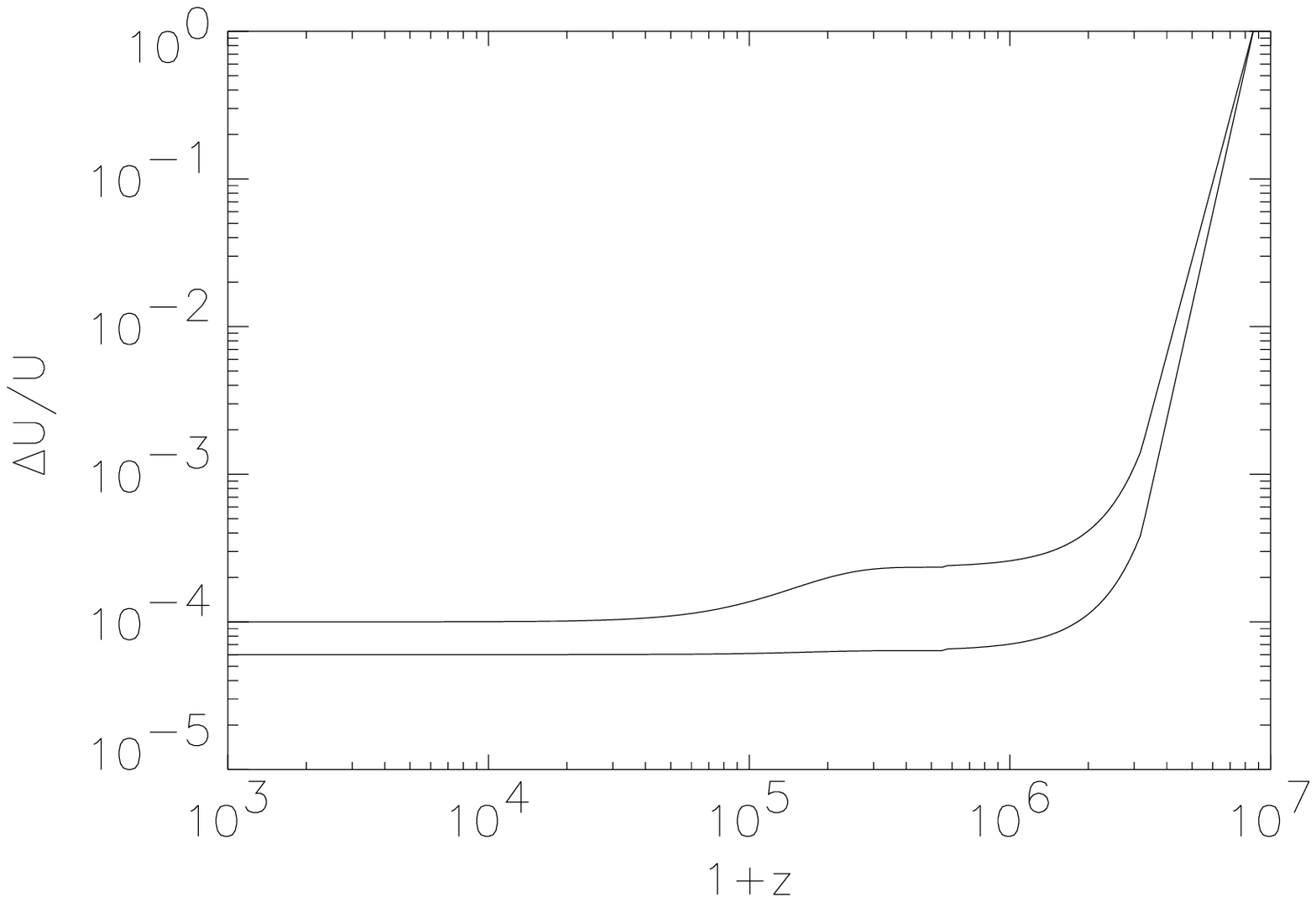}}
\begin{description}
\item[Figure 4.] Limits on energy injection into the CMB
prior to decoupling. The top curve uses $y=2.5\times 10^{-5}$
and $\mu = 3.3\times 10^{-4}$. These were the limits from FIRAS in 1994
(Wright {\it et al.} \cite{Wright94}). The different
epochs are clearly evident. The bottom curve is based on
the FIRAS limits in eq. 6.
This plot was produced with a program written by Ned Wright
and is based on models in Burigana {\it et al.} \cite{Burigana91a},
\cite{Burigana91b}.
\end{description}
\end{figure}

	The $y$ distortion is manifest at high frequencies and it
will be a long time before the FIRAS limit is improved. The limit
on $y$ also strongly constrains alternative models of the origin of
the CMB. One may try to mimic a Planck spectrum with a
superposition of multiple grey bodies. At long wavelengths, the
Rayleigh-Jeans region, the results cannot be distinguished.
However, near the  peak of the spectrum, such a superposition
will result in a $y$-distortion. We also
note that if the universe is inhomogenous on the largest scales,
and we are not at a preferred center, a distortion will result
\cite{Nature}.

	The signatures of any $\mu$ and $Y_{ff}$ distortions are evident at
low frequencies. While the current generation of experiments will
just barely, if at all, improve on the FIRAS  limits, they are
paving the way for the next generation  which may
detect a distortion. Figure 5 shows a plot of the spectrum along
with the $\mu$ and $Y_{ff}$ distortion limits.

	Two groups \cite{Staggs96}, \cite{Shafer96} are pursuing
long-wavelength measurements of the spectrum. Outside of the
precise instrumentation necessary to  perform absolute
measurements between 0.1\% and 1\%, one must contend with
Galactic and atmospheric emission. At 600 MHz, the temperature of
the Galaxy is
roughly 5.8 K \cite{Smith96} and falls as
$\nu^{-2.7}$.  In Figure 1, this is roughly where the
``synchrotron'' and ``CMB'' lines cross.  Observing from the ground, the
atmosphere emits at roughly 1.5 K between 0.6 GHz and 1.4 GHz
\cite{Smith96}, \cite{Staggs94}. Ground-based experiments thus
require precise modeling and sky dips to subtract the
atmospheric signal. Compact experiments may be flown from
balloons to rise above it. Finally, in the not-too-distant
future, narrow-band measurements will be limited by the FIRAS
error when characterizing distortions. 

\begin{figure}
\epsfysize=3.5truein
\epsfverbosetrue
\centerline{\epsfbox{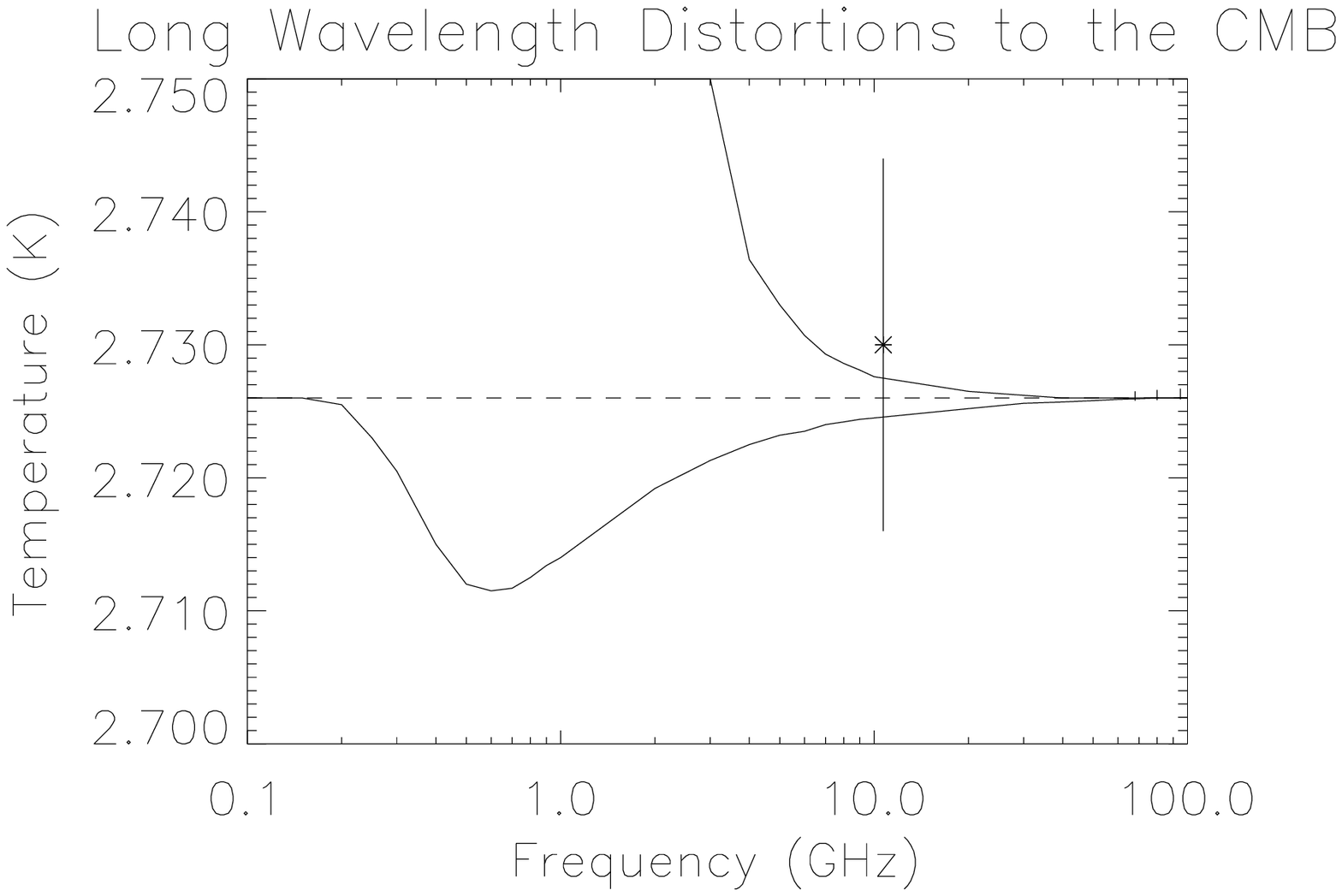}}
\begin{description}
\item[Figure 5.] Plot of the long wavelength (low
frequency) distortions  to the CMB. The flat line is for a
perfect blackbody. The curve with a pronounced minimum near 700
MHz shows a $\mu$ distortion with $\mu = 9\times10^{-5}$, the
{\sl COBE} limit. The curve that begins to rise near 10 GHz is
for a $Y_{ff}$ distortion with $Y_{ff} = 1.5\times 10^{-5}$. All
data with error bars small enough to fit on this plot are shown.
The measurement at 10.7 GHz comes from Staggs {\it et
al.} \cite{Staggs96}. Note the three FIRAS data points near 100
GHz. This plot was adapted from a similar plot made by Al Kogut.  
\end{description}
\end{figure}

\section{The Anisotropy} 

The photons that end their lives in our detectors were last
scattered off electrons when the universe had a temperature of
5000 K and was evolving from a plasma to a state of neutral
hydrogen and helium plus a thermal background\footnote{We assume
the standard inflationary model in this discussion, as well as a
nearly complete transition from plasma to neutral matter.}. This
era is called the epoch of decoupling. The plasma  was responding
to the gravitational potential wells that would eventually foster
galaxies and clusters of galaxies. The photons, now decoupled,
bring to us an imprint of the  potential wells and a signature of
the dynamics of the plasma's response to the wells. From the
angular spectrum of the fluctuations, one can distinguish among
various possible mechanisms of structure formation. In recent
years,  it has become evident that for a certain class of models
(eg. standard CDM\footnote{Standard CDM has $\Omega_0 =1$, $h=0.5$,
$\Omega_B=0.05$, and $n=1$. Ratra points out that ``fiducial''
is a better description because many observations disagree with 
standard CDM.}), a measurement of the detailed shape of the power spectrum
will yield values for many of the cosmological parameters
\footnote{$\Omega$ is the fraction of the critical density. For
standard Cold Dark Matter models, $\Omega_{0} = \Omega_{CDM} +
\Omega_{B}$; for Lambda models,  $\Omega_0 + \Omega_{\Lambda}=1$;
while for open models,  $\Omega_{0} + 
\Omega_{\Lambda} + \Omega_{curv} = 1$.}such as $\Omega_0$,
$\Omega_B$, $H_0$, and $\Lambda$ \cite{Jungman96}.

	Figure 6 shows the angular spectra for a few of the many
models of structure formation. The point of the plot is to
indicate that the model predictions are rather different and that
given measurements with uncertainties of order the width of
the plot line, the best model could be identified.  

\begin{figure}
\epsfysize=3.5truein
\epsfverbosetrue
\centerline{\epsfbox{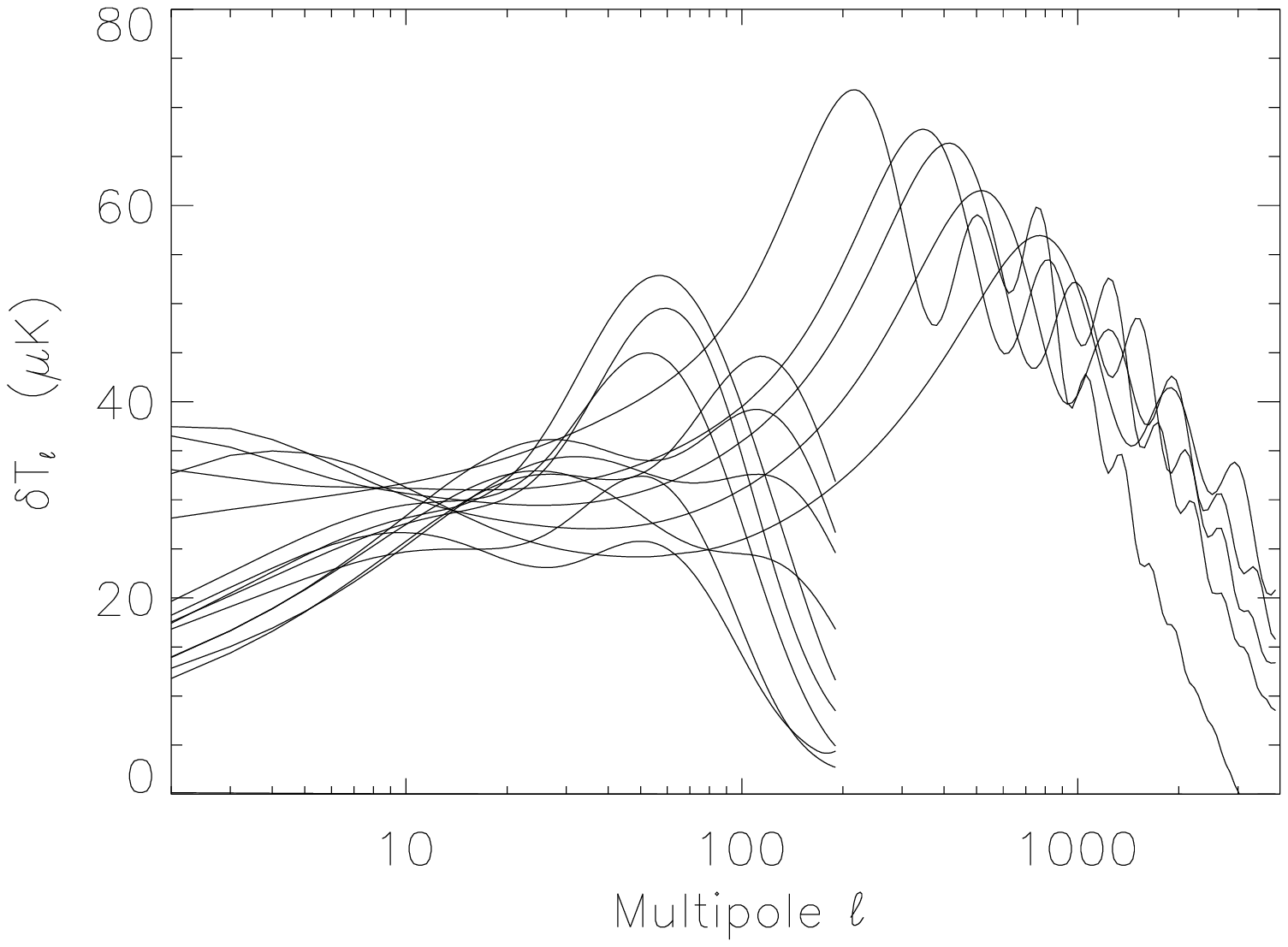}}
\begin{description}
\item[Figure 6.] Angular spectrum for a sample of anisotropy models.
The curves grouped on the left hand side are the predictions 
of isocurvature models from Peebles \cite{Peebles96}. The curves that 
peak on the right are for the standard and open
bubble inflation models (Ratra and Sugiyama \cite{Rat95a}, [19]) with 
$\Omega_0 = 1.0, 0.4, 0.3, 0.2$ and $0.1$ for peaks going from left to right.  
\end{description}
\end{figure}

	We divide the spectrum into three regions. At $l<80$, or
large angular scales ($\delta\theta>2^{\circ}$, $l\approx
180/\delta\theta$ with $\delta\theta$ in degrees), separate
regions of the sky are not causally connected at decoupling.
Indeed, the relative isotropy at these scales was one of the
motivations for the  inflation model.  If inflation is correct,
these  potential wells and hills are the manifestation of quantum
fluctuations that were superluminally expanded beyond the Hubble
radius, grew with the expansion of the universe, and then
re-entered  the Hubble radius at a later time. The largest scales
entered the Hubble radius most recently. In 10 billion years, a
new CMB quadrupole will come into view as the universe expands
and what we now call the quadrupole will be distributed among the
higher moments. At these large scales, the anisotropy is produced
by photons climbing out of the gravitational potential wells or
sliding down the hills, just after their last scattering event, as
discovered by Sachs and Wolfe. The anisotropy in this region
reflects the primordial power spectrum of fluctuations, P(k). One
of the strengths of inflation is that it predicts the shape of
this spectrum\footnote{The shape  of the spectrum was surmised
independently by Harrison, Peebles \& Yu, and Zel'dovich long before
inflation.}. At large angular scales, $C_l\propto 1/l(l+1)$, in
other words, a  nearly flat line in Figure 6.

	At smaller angular scales, greater than $l\approx 80$,  there
was time for the primordial plasma to communicate. Hu and
colleagues \cite{HSS96}, \cite{Tegmark95} have presented an
intuitive physical picture of the mechanisms behind the
anisotropy although
models date back to Silk \cite{Silk68}, 
Sunyaev and Zel'dovich \cite{SZ70}, Peebles \& Yu
\cite{Peebles70} and others. To first order, we may think of the plasma as
a photon-baryon fluid that acoustically oscillates
in response to fluctuations
in the gravitational potential produced by the dark matter. As
the fluid flows into a potential well, it heats up (the phase is
opposite to  that of the Sachs-Wolfe effect). This is responsible
for the large peak in the power spectrum which occurs at roughly
the angular scale of the largest potential that can support
plasma oscillations. Oscillations of the 
fluid in response to fluctuations at smaller scales  give rise to the
other peaks. A full description accounts for the doppler effect,
the self gravity of the  photon-baryon fluid, the inertia of the
baryons, and the evolution of the fluid with time. 

	Because decoupling happened so fast, $\delta z/z \approx
0.1$, the thickness of the surface of last scattering is less
than the horizon size at $z\approx 1400$. The scale of the
fluctuations near the first peak is of order the horizon size,
about 100 Mpc in comoving units\footnote{Alex Szalay spoke of a
possible  connection between this scale and the scale of the
largest structure in the galaxy surveys.}, and we observe the
full signature of the potential wells there. On smaller angular
scales, corresponding to smaller physical scales, there are more
fluctuations contained within the horizon and their effects
average out. In addition, diffusion of photons out of the
potential wells diminishes the temperature fluctuations. The
combination of these effects leads to the suppressed anisotropy
near $l=2000$ as can be seen in Figure 6.

	Cosmological parameters are extracted from the shape of
the power spectrum. For standard CDM  type models this is done,
for instance, by noting that the distance
between the peaks depends on the sound speed at decoupling. This
in turn depends on $h^2\Omega_B$. Also, as $\Omega_B$ increases,
the photon-baryon fluid has more inertia (the sound speed
decreases) and the compressional peaks (the odd ones starting at
$l=200$)  get bigger. On the other hand, if the universe goes
through a period of reionization, all the peaks in the anisotropy
can be wiped out. Extracting generic cosmological information,
regardless of model, is an active area of research \cite{HW96}.

	One particularly nice demonstration of what the 
anisotropy can tell us was noted by Kamionkowski {\it et al.}
\cite{Kamionkowski94}. The location in $l$ of the first peak is a
good indicator of $\Omega_0 +\Omega_{Lambda}$.  This happens
because a universe of any geometry, in its early stages, evolves
as though $\Omega_0 = 1$. Because the physical size of a
fluctuation depends on the  sound horizon at decoupling, it is a
``standard yardstick.'' The  angular size of the standard
yardstick, as viewed today, depends on the overall geometry of
the universe. If the universe is flat, it will appear at
$2^{\circ}$, if it is open it will appear at  a smaller angle;
this last is because there is ``more space'' far away. This
effect is seen in the CDM models in Figure 6. The lower
$\Omega_0$, the further to the right the peak moves. The scaling
is roughly $l_{peak} \approx 200/\Omega_0^{1/2}$.

	In models of the formation of large scale  structure, the
same potential fluctuations that produce the anisotropy also
produce the large scale clustering of galaxies. We still do not
know what type of mass comprises galaxies, or how it couples to
gravitational fluctuations, or even what produced the
fluctuations. In the past few years, a number of large galaxy
surveys have been undertaken and  older surveys have been
re-analyzed. Over a region of $l$-space between $l = 30$ and $l =
600$, there is overlap between the two probes of the 
fluctuations: galactic surveys and CMB anisotropy data. There is
a nice plot of this in White, Scott, \& Silk's review
\cite{WSS94}.  Unfortunately, the data are not sufficiently good,
from either probe, to draw firm conclusions. However, the
indications are that the simplest models for the formation of
structure are incorrect.       

\section{Anisotropy Measurements: Technologies \& Techniques }

\subsection{Technologies}

	The desire to characterize the anisotropy has led to 
improvements in detectors and instrument technology. For the
anisotropy, three classes of detector are now in use:  HEMT based
amplifiers (20-90 GHz), superconductor-insulator-superconductor
(SIS) based mixers (90-250 GHz), and a variety of types of
bolometers (90-1000 GHz).  Anisotropy measurements require very
stable observing conditions. This is especially true for
``configuration-space'' measurements, as opposed to
interferometric measurements. To overcome atmospheric
fluctuations, the primary culprit, experiments are performed at
stable or high sites (Mauna Kea, South Pole, Owens Valley,
Saskatoon, \& Tenerife) and from balloon platforms. Plans are
underway for long duration balloon flights that circumnavigate the
Antarctic and for balloons that can stay aloft for 100 days.
Interferometers, which currently operate mostly below 30 GHz, are
intrinsically less sensitive to the atmosphere. 

	The HEMT amplifiers for most of the  HEMT-based
experiments were designed by Marian Pospieszalski at the National 
Radio Astronomy Observatory (NRAO) electronic research lab. One
couples celestial radiation to the amplifiers with waveguide. The
incident electric field is simply amplified to a reasonable level
and then detected with a diode. The amplifiers are special
because they are rugged, require only modest cooling ($\approx
20$ K), and have a 30\% bandwidth. A good wide-band (10
GHz) sensitivity for an optimized system near 40 GHz is roughly  ${\rm
500~\mu Ks^{1/2}}$. In other words, one can detect  1/2 mK signal
with a signal-to-noise of one with one second of integration. A
number of  semi-conductor groups are pushing to make better
high-frequency transistors (for instance TRW and Hughes) and to
make entire radiometers on a single chip (TRW and
Lockheed-Martin). 
  
	SIS-based systems have been used for a number of years
(Timbie \cite{Timbie89}, Meinhold \cite{Meinhold90},  Robertson
\cite{Robertson96}) though the anisotropy has not yet been
detected with them. The currently favored designs, and devices,
come from Anthony Kerr and S-K Pan at NRAO. The SIS  is a mixer.
It is simultaneously illuminated with  celestial radiation and
with a ``local oscillator.'' The output of the SIS is a signal
containing the sum and difference frequencies of the LO and sky.
The signal is low-pass filtered, amplified with HEMTs, and
detected. For a system with a LO at 144 GHz and a 4 GHz IF
bandwidth, a reasonable sensitivity is  ${\rm 400~\mu Ks^{1/2}}$.
These devices must operate below the superconducting transition
of niobium, or below roughly 4.2 K. They are more difficult than
HEMTs to operate but are still straightforward.        

	Both HEMT and SIS systems are coherent; in other words,
one works with the electric field right up until the final
detection. This allows possibilities for phase sensitive
techniques such as correlation receivers and interferometers. In
addition they both have a very large audio bandwidth. Signals at 
many megahertz are easily detected. 

	The  current bolometers are essentially thermistors held
at 0.3 K or below. When they are illuminated, they heat up and
change resistance, and the change in resistance is electronically
read out. This is called incoherent detection because the phase
information is lost. The great advantage to bolometers is that
they are very sensitive. A decade ago, they  were near ${\rm
400~\mu Ks^{1/2}}$ \cite{FIRS}, and  some current devices (the
``spider'' bolometers, Bock et al. \cite{Bock96}) achieve better
than  ${\rm 100~\mu Ks^{1/2}}$. The disadvantage is that they are
more difficult to use than HEMTs and SISs and their intrinsic
time constants are longer,  but both of these problems are
actively being worked on. There are different types of bolometers
in various stages of development. They include frequency
sensitive bolometers\cite{Kowitt96}, hot electron
bolometers\cite{Nahum93}, monolithic silicon
bolometers\cite{Downey84}, spider web composite bolometers\cite{Bock96},  
and transition edge bolometers\cite{Lee96}.  

	The bolometer's advantage is not its intrinsic
sensitivity per photon, which is comparable to that of HEMTs or
SIS, but rather the fact that one can detect in multiple
electromagnetic modes (in conventional use, waveguide supports
just one mode) and almost arbitrary bandwidth. In other words,
bolometers detect more photons than single mode waveguide
systems. The power on a device is 

\begin{equation} 
P= \int_{\nu} \int_{A} \int_{\Omega} S_{\nu}(T) d\nu dA d\Omega \rightarrow 
\int_\nu kT(\nu)d\nu~~~{\rm Watts},
\end{equation}
where $S_\nu(T)$ is the flux from some source, $\nu$ is the RF
frequency, $A$ is the area of the detector (or antenna), and
$\Omega$ is the solid angle incident on that detector (or
antenna). The quantity  $\int dA d\Omega$ is called the
throughput or \'etendue. Generally, $A\Omega = n\lambda^2$ where 
$\lambda$ is the wavelength at the passband center and $n$ is the
number of modes. For a single mode system, $n=1$, and we get
the quantity on the right of the arrow, where $T(\nu)$ is the effective
temperature of the source. 

\subsection{Techniques}

	The experimental challenge is to measure a variance of a 
random field, which is of order 30 $\mu $K, with a noisy detector and
a background signal of 300 K. As much effort has gone into 
determining robust ways to do this as has gone into understanding
the detector systems. The configuration-space techniques are the
better developed so we will focus on those. The interferometric
techniques are rapidly maturing and have a lot of promise; the first
detection of the anisotropy with an interferometer was just
reported in Scott {\it et al.} \cite{CAT}. 

	A typical telescope has a beam described by  $P(\theta)
\propto \exp(-\theta^2/2\sigma^2)$. If the telescope observes the
sky which has temperature $T(\hat x)$, it measures $t(\hat x) =
\int P(\theta )T(\hat x)dx $. If two measurements made near each
other, $t(\hat x_1)$ and $t(\hat x_2)$ are subtracted, the common
atmospheric signal drops out.  One is then sensitive to only the
gradient in the atmosphere and to the CMB temperature difference.
If two single differences, with one position in common,  are
performed  and then subtracted the ``double difference'' is  $s_i
= 2t(\hat x_1) - t(\hat x_2) - t(\hat x_3)$.  The double
difference is still sensitive to the  CMB but is insensitive to
atmospheric gradients as well as the atmospheric temperature.
This extra difference helps because the fluctuation spectrum of
the atmosphere drops with both increasing spatial and spectral
frequency. Let us call the  effective beam profile  for the
double difference measurement $H_i(\hat x)$; it will have one
large positive central lobe and two negative lobes as shown in
Figure 7. We can write
     
\begin{equation} 
s_i = \int H_i(\hat x)T(\hat x) dx.
\end{equation}

We will consider this a single measurement at a pixel $\hat x$. In practice, 
we make a set of $N$ similar measurements over a patch of sky.
For a first order estimate, we may find the intrinsic variance of
the sky from
\begin{equation} 
\sigma^2_{sky} = \sigma^2_{meas} - \sigma^2_{data},
\end{equation}
where $\sigma^2_{data}$ is the square of the average statistical error per
measurement and $\sigma_{meas}$ is the variance of the $N$ data 
points. This answer is usually only correct to 25\% so it is used 
as a sanity check. Also, it does not give the correct error. 
Note too that this method ignores all intrinsic correlations
in the data.

What variance does one expect from such a measurement? After
working through the math, we find that the inclusion of
finite beams modifies eq.\ 3 to the following:
\begin{equation} 
C^{ij}_T = \sum_L{2l+1\over 4\pi} C_l W_l^{ij},~~
{\rm where}~~W_{l}^{ij} = 
\int dx_1 \int dx_2 H_i(\hat x_1)H_j(\hat x_2) P_l(\hat x_1\cdot
\hat x_2).
\end{equation}
This is the full theoretical covariance matrix for the observing
pattern for measurements $i$ and $j$.\footnote{From a theoretical
perspective, our universe is one realization of a
cosmological model that can only predict the ensemble average of
$C_l$. Even if we knew the correct physics, the data would
be scattered around the predicted $C_l$ with a ``cosmic variance.''}
$W_l$ is called the window function, it tells us the
portion of $l$-space being examined. Generally, just the  diagonal
elements are plotted. For the beam in Figure 7,  the
window function is shown in Figure 8. When $i=j$, we get a
prediction for $\sigma^2_{sky}$.

\begin{figure}
\epsfysize=3.8truein
\epsfverbosetrue
\centerline{\epsfbox{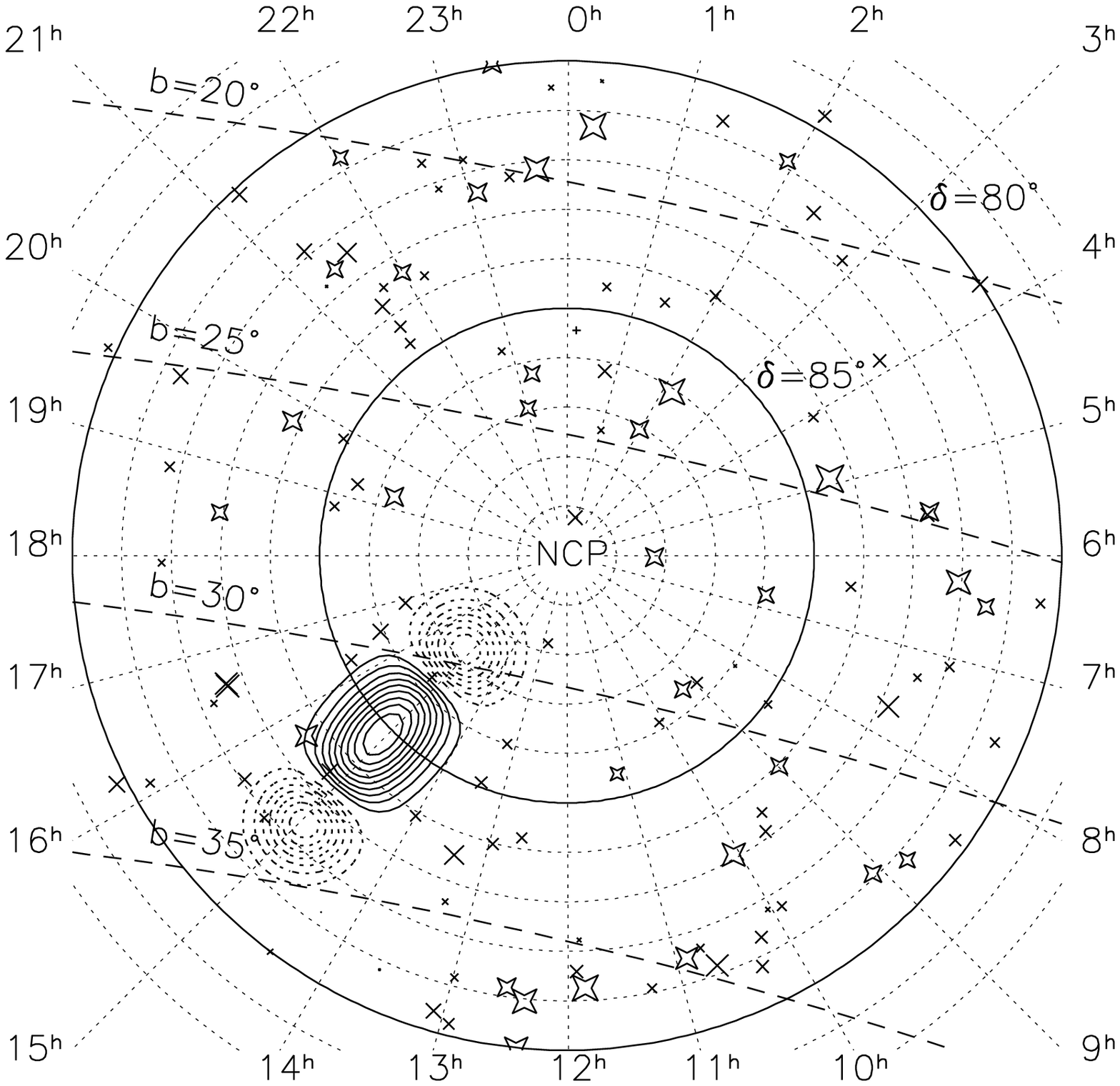}}
\begin{description}
\item[Figure 7.] Contour plot of a typical double-difference
beam superimposed on a source map of the north celestial polar
region from Netterfield {\it et al.} \cite{Netterfield95}.
This profile corresponds to $H_i(\hat x)$ in eq.\ 8.
The dashed lines are negative and the solid lines are
positive. The sources come from the K\"uhr survey\cite{Kuhr81}. 
The stars mark the flat spectrum sources. The symbol size is
proportional to the log of the flux. Lines of Galactic latitude
are also shown.  
\end{description}
\end{figure}

\begin{figure}
\epsfysize=3.truein
\epsfverbosetrue
\centerline{\epsfbox{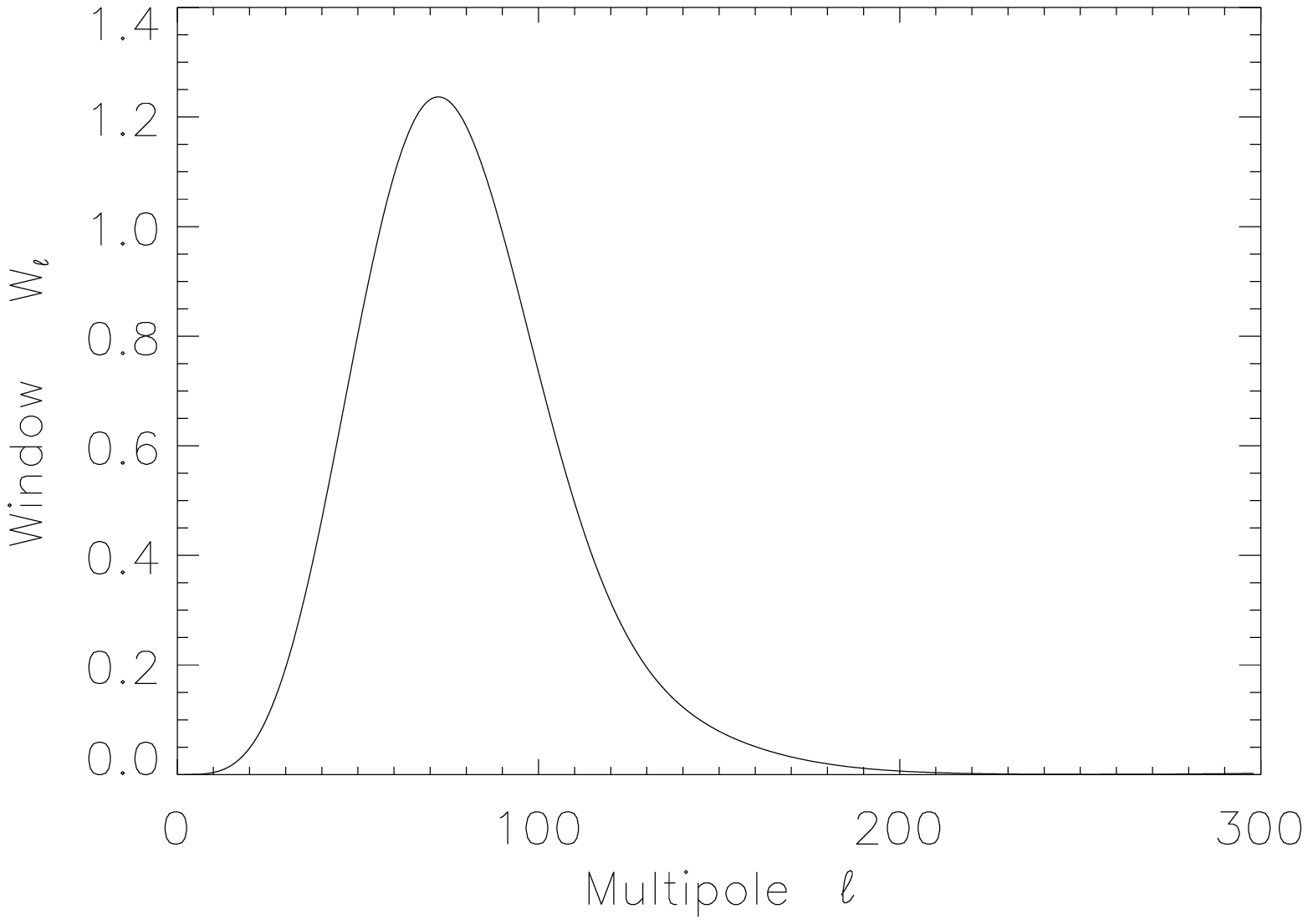}}
\begin{description}
\item[Figure 8.] Window function for the beam in Figure 7
using eq.\ 10 with $i=j$. The
amplitude of the window depends on the 
normalization of the beam. For this plot, $\int d\hat x|H(\hat x)| = 2$
was used.
\end{description}
\end{figure}

	The most frequently used analyses follow Bond's
work \cite{Bond95}.
A complete analysis requires knowledge of the 
covariance matrix of the data; we call this
$C_D$. The full theory-plus-data covariance matrix is
given by $M = C_D + C_T$. Our goal is to determine
the probability of a theory given the data, $P(T|D)$. To do this,
we use Bayes's theorem with a uniform prior, P(T) = 1, and 
set the probability of getting the data, P(D), equal to 1:

\begin{equation} 
P(T|D) = {P(D|T)P(T)\over P(D)} = L(D|T) = {\exp(-t^{T}M^{-1}t/2)\over
(2\pi)^{N/2}|M|^{1/2}},
\end{equation}
where L is the likelihood function and $t$ is a vector 
of the data. The argument of the exponent is proportional
to $\chi^2$. This boils down to saying that the likelihood  of
the data plotted as a function of some parametrization, for
instance $\sigma_{sky}$, is the probability of obtaining
$\sigma_{sky}$ with a given set of data. When the signal to noise
is high, the likelihood is fairly Gaussian. To  get an error,
we find bounds symmetric around the maximum, that contain 68\% of
the area under the likelihood curve.

	To estimate the angular spectrum, we take the most likely
value of $\sigma_{sky}$ with its error, convert it into a
``band-temperature''\footnote{See Bond \cite{Bond94} for details.  One
obtains a band power by dividing $\sigma_{sky}$ by $\sqrt{\sum_l
W_l/l}$. This makes sense because $\sigma_{sky}=\sqrt{\sum(\delta
T_l)^2W_l/l}$ can be written as $\sigma_{sky}=\bar{\delta
T_l}\sqrt{\sum W_l/l}$.} and plot it at the $l$-weighted center
of the  window function. A horizontal error bar is often plotted
indicating the width of the window function. Any one experiment
observes in multiple windows and so the spectrum may be mapped
out. One example is given in Figure 9 from the Saskatoon
experiment \cite{Netterfield97}. 

\begin{figure}
\epsfysize=3.5truein
\epsfverbosetrue
\centerline{\epsfbox{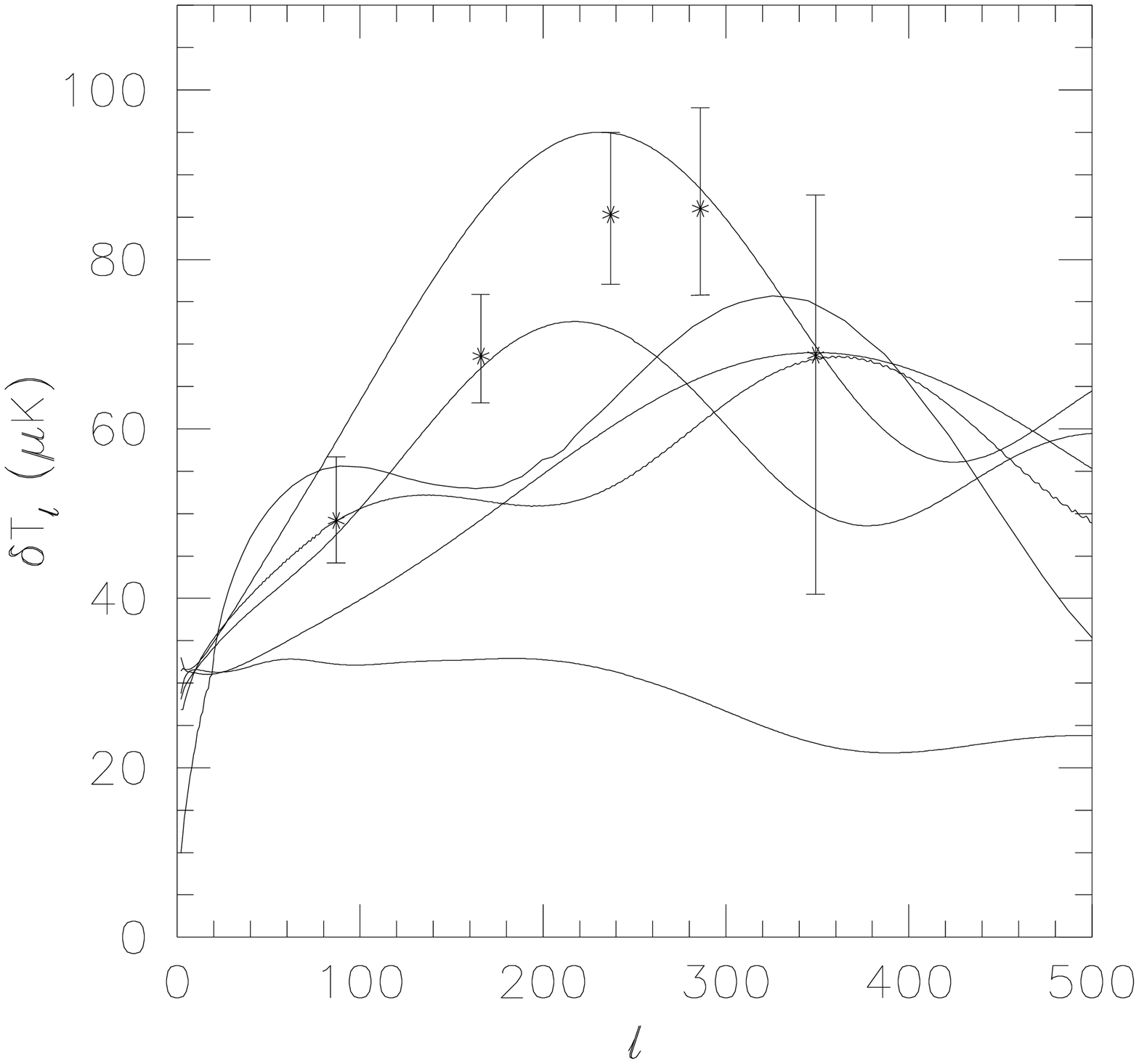}}
\begin{description}
\item[Figure 9.] Results from the Saskatoon experiment 
\cite{Wollack97}, \cite{Netterfield97}. The spectra from six theories are also
shown. From top to bottom at $l=160$ they are a flat
$\Lambda$+CDM model with $\Omega_\Lambda=0.7$ \cite{Rat95a},
Standard CDM \cite{Rat95b}, a PPI model \cite{Peebles95}, an 
$\Omega_0=0.4$ open bubble model \cite{Rat95b}, a texture
model \cite{Cri95}, and a model with reionization
\cite{Sug95}. There is a  $14\%$ overall calibration
uncertainty that is not included in the error bars. This affects
the normalization of the spectrum, but not the shape.
\end{description}
\end{figure}

\section{Anisotropy Measurements:\protect\\ 
The Current Results and Immediate Future }

	The anisotropy of the CMB was first unambiguously
measured by the DMR experiment aboard the {\sl COBE} satellite
\cite{Smoot92} using $7^{\circ}$ resolution full-sky maps at 30,
53, and 90 GHz.  To date, these are still the cleanest and best
checked data. All indications are that the fluctuations are
thermal, though I am not aware of any formal limits on, say, the
Compton $y$ parameter of the anisotropy. When smoothed to a
$10^{\circ}$ resolution, the {\it rms} of the temperature
fluctuations is about $30~\mu$K. This is the canonically quoted
value. However, with a $1/2^{\circ}$ resolution map, the {\it
rms} is closer to $90~\mu$K. The final DMR results (the satellite
is now turned off) are published in Ap.J Vol 464, 1996 
\cite{Bennett96}, \cite{Kogut96}, \cite{Gorski96},
\cite{Hinshaw96}, \& \cite{Wright96}. The lasting contribution
will be the maps of the sky at 30, 53, and 90 GHz. A combination
of these maps, optimized to give the anisotropy, has a
signal-to-noise of two per $10^{\circ}\times 10^{\circ}$ pixel.
The two primary results derived from these maps are:

\begin{enumerate}

\item From a fit of the data to a power spectrum parameterized by
the spatial index and the quadrupole amplitude, the DMR
team finds $n_{DMR} = 1.21\pm 0.3$ and $Q_{rms-PS} =
15.3^{+3.8}_{-2.8}~\mu$K. The quadrupole of the raw maps is
slightly below $Q_{rms-PS}$, but not by a statistically
significant amount. Note that for ``standard CDM'' one expects
$n_{DMR} = 1.1$ because DMR probes the low-$l$ tail of the
acoustic peak. In this notation, the $C_l$ (eq.\ 10) are given
by \cite{BE87}: 
\begin{equation} 
C_l={4\pi\over 5}Q^2_{rms-PS}
{ \Gamma [l+(n_{DMR}-1)/2]\Gamma [(9-n_{DMR})/2]\over 
\Gamma [l+(5-n_{DMR})/2]\Gamma [(3+n_{DMR})/2 }
\end{equation}
with $n_{DMR}=1$ this reduces to $C_l\propto 1/l(l+1)$.

\item The data appear best described by Gaussian statistics
\cite{Kogut96b}. At these
large angular scales this is not surprising because even
non-Gaussian processes at small scales, when averaged over a
large enough  volume, appear Gaussian. However, it is reassuring
that the statistics we all assume have some basis in reality.

\end{enumerate}

	My favorite way of quantifying the DMR maps is to show
the power spectrum over compact regions of $l$-space. A direct
computation  of the power spectrum is hampered by the
unevenly-weighted non-uniform sky coverage imposed by the scan
pattern and the elimination of data contaminated by Galactic
emission. There are a number of ways around this problem 
(\cite{HP73}, \cite{Gorski94}). The most recent approaches
(\cite{Tegmark96}, \cite{Gorski96b}) work to minimize the width of
a representative bin in $l$-space. The results are shown in
Figure 10. G\'orski's method in particular shows the power within
$\Delta l=0$.  Though one should expect to get exactly the same
results from two different methods applied to the same data set,
we can be pleased by the general concordance.

\begin{figure}
\epsfysize=3.5truein
\epsfverbosetrue
\centerline{\epsfbox{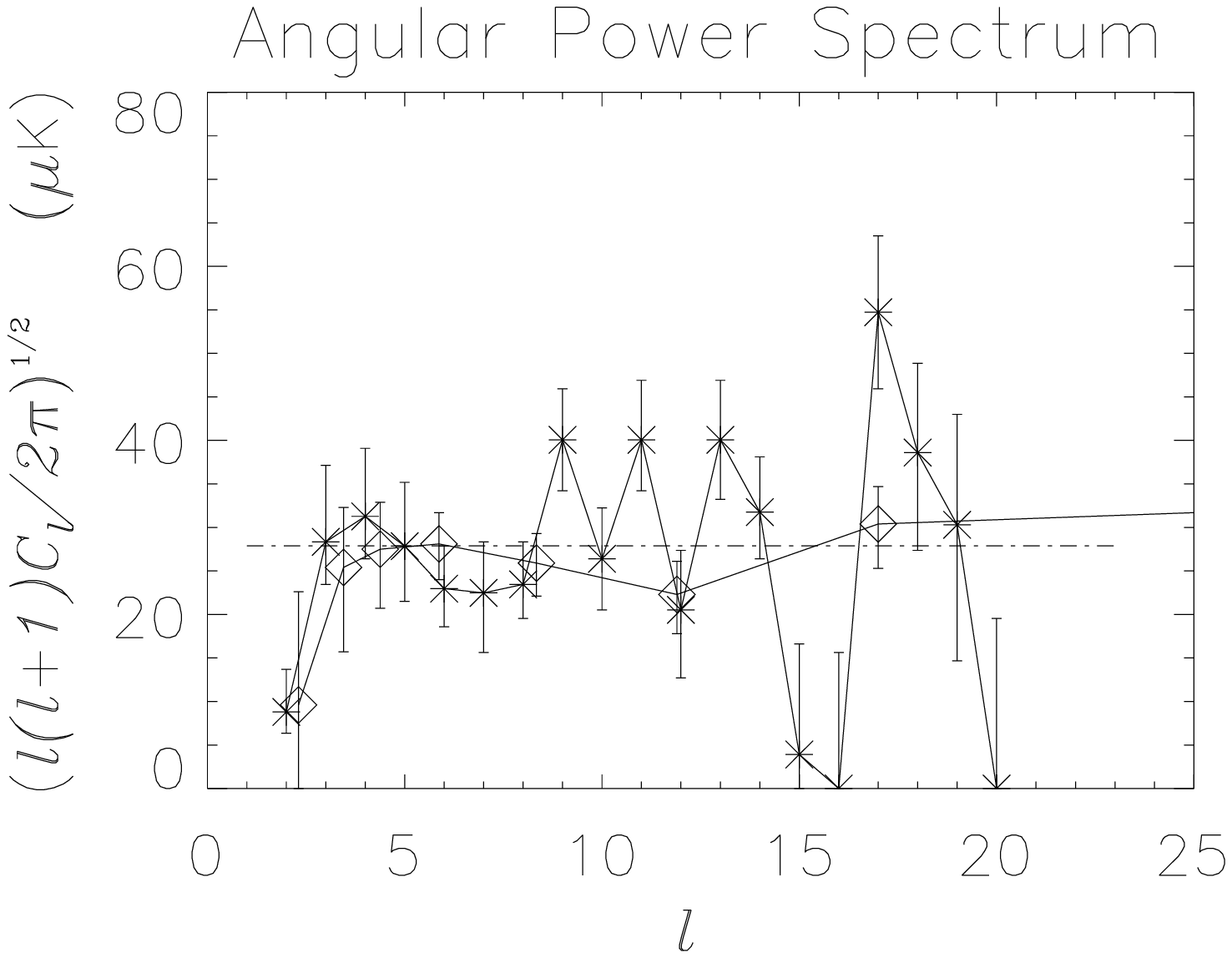}}
\begin{description}
\item[Figure 10.] The angular power spectrum from DMR. Results from
G\'orski \cite{Gorski96b} (stars) and Tegmark \cite{Tegmark96}
(diamonds) are shown. Both have analyzed the data in a manner to
produce narrow window functions. G\'orski's, in particular, have
$\Delta l/l = 0$. One should bear in mind that G\'orski's results
are derived from non-Gaussian likelihood distributions.
The flat dashed line is for a Harrison-Zel'dovich
spectrum with $Q_{rms-PS}=18~\mu$K.  
\end{description}
\end{figure}

	Many groups are working to measure the
anisotropy. Though some are focussing on large angular scales and
frequencies not observed with DMR, most concentrate on smaller
angular scales. Table 1 contains a list for recent, current and
planned experiments. It does not include the satellite experiments
nor does it claim to be comprehensive. I apologize for any
omissions or misrepresentations.

\begin{table}[p]
\begin{center}
\caption{Recently Completed, Current and Planned Anisotropy
Experiments}
\begin{tabular}{lccccl}
\hline
Experiment & Resolution & Frequency & Detectors & Type & Groups\\
\hline
ACE(c)\cite{ACE}               & $0.2^{\circ}$ & 25-100 GHz & HEMT & C/B  & UCSB \\
APACHE(c)\cite{APACHE}       & $0.33^{\circ}$ & 90-400 GHz & Bol & C/G  & Bologna, Bartol \\
                                                 &&&&& Rome III \\
ARGO(f)\cite{ARGO}  & $0.9^{\circ}$ & 140-3000 GHz & Bol & C/B  & Rome I \\
ATCA\cite{ATCA}    & $0.03^{\circ}$ & 8.7 GHz & HEMT & I/G & CSIRO \\
BAM(c)\cite{BAM}     & $0.75^{\circ}$ & 90-300 GHz & Bol & C/B & UBC, CfA \\ 
Bartol(c)\cite{TR2} & $2.4^{\circ}$  & 90-270 GHz  & Bol & C/G & Bartol \\ 
BEAST(p)\cite{ACE}                 & $0.2^{\circ}$ & 25-100 GHz & HEMT & C/B  & UCSB \\
BOOMERanG(p)\cite{BOOMERanG}       & $0.2^{\circ}$ & 90-400 GHz & Bol & C/G 
                                                 & Rome I, Caltech\\
                                                 &&&&&  UCB, UCSB \\
CAT(c)\cite{CAT}         & $0.17^{\circ}$ & 15 GHz & HEMT & I/G & Cambridge \\
CBI(p)\cite{CBI}               & $0.0833^{\circ}$  & 26-36 GHz & HEMT & I/G & Caltech, Penn. \\
FIRS(f)\cite{FIRS}   & $3.8^{\circ}$ & 170-680 GHz & Bol & C/B & Chicago, MIT,\\ 
                                                  &&&&&Princeton,\\
                                                  &&&&&NASA/GSFC \\
HACME/SP(f)\cite{HACME}    & $0.6^{\circ}$  & 30 GHz & HEMT & C/G & UCSB \\
IAB(f)\cite{IAB}    & $0.83^{\circ}$ & 150 GHz & Bol & C/G & Bartol \\
MAT(p)\cite{MAT}    & $0.2^{\circ}$  & 30-150 GHz & HEMT/SIS &
C/G & Penn, Princeton \\
MAX(f)\cite{MAX}    & $0.5^{\circ}$  & 90-420 GHz & Bol & C/B & UCB, UCSB \\
MAXIMA(p)\cite{MAXIMA}   & $0.2^{\circ}$  & 90-420 GHz & Bol &
C/B & UCB, Caltech \\
MSAM(c)\cite{MSAM}    & $0.4^{\circ}$  & 40-680 GHz & Bol & C/B & Chicago, Brown, \\
                                                &&&&& Princeton, \\
                                                &&&&& NASA/GSFC \\
OVRO 40/5(c)\cite{OVRO} & $0.033^{\circ},0.12^{\circ}$ & 15-35 GHz & HEMT & C/G & Caltech, Penn \\ 
PYTHON(c)\cite{PYTHON} & $0.75^{\circ}$ & 35-90 GHz & Bol/HEMT & C/G & Carnegie Mellon\\
						&&&&& Chicago, UCSB\\ 
QMAP(f)\cite{QMAP} & $0.2^{\circ}$ & 20-150 GHz  & HEMT/SIS & C/B & Princeton, Penn \\ 
SASK(f)\cite{SASK} & $0.5^{\circ}$ & 20-45 GHz  & HEMT & C/G & Princeton \\ 
SuZIE(c)\cite{SuZie} & $0.017^{\circ}$ & 150-300 GHz  & Bol & C/G & Caltech \\ 
TopHat(p)\cite{TopHat} & $0.33^{\circ}$ & 150-700 GHz & Bol & C/B & Bartol, Brown,\\ 
                                       &&&&& DSRI,Chicago, \\
                                       &&&&& NASA/GSFC\\
Tenerife(c)\cite{Tenerife}  & $6.0^{\circ}$  & 10-33 GHz  & HEMT & C/G & 
NRAL, Cambridge \\ 
VCA(p)\cite{VCA} & $0.33^{\circ}$ & 30 GHz  & HEMT & I/G & Chicago \\ 
VLA(c)\cite{VLA} & $0.0028^{\circ}$ & 8.4 GHz  & HEMT & I/G & Haverford, NRAO \\ 
VSA(p)\cite{VSA} & -- & 30 GHz  & HEMT & I/G & Cambridge \\ 
White Dish(f)\cite{WD} & $0.2^{\circ}$ & 90 GHz  & Bol & C/G & Carnegie Mellon \\ 
\hline
\end{tabular}
\end{center}
\begin{enumerate}
\item For ``Type'' the first letter distinguishes between configuration
or interferometer, the second between ground or balloon.
\item An ``f'' after the experiment's name means it's finished;
a ``c'' denotes current; a ``p'' denotes planned, building may 
be in progress but there is no data yet.
\end{enumerate}
\end{table}

	Unlike measurements of the absolute temperature of the CMB,
where the final result is completely dominated by one's 
control of subtle systematic errors, anisotropy measurements require
a combination of high sensitivity and immunity to systematic
effects. The state-of-the-art in absolute measurements,
excluding FIRAS, is 1\% \cite{Staggs96}; the anisotropy 
has yet to be measured to 15\% accuracy.

To give a broad and almost un-biased sense of what the
anisotropy  data are telling us, we take the compilation from
Ratra \cite{Ratra96}
(which I believe is the most comprehensive and thoroughly checked 
compilation to date) and bin the data according to the following:
\begin{itemize}
\item Select logarithmically spaced bins in $l$ with four bins per decade.
\item Ignore the widths of window functions and add data to a
bin according to the value of the weighted mean of the window,
$l_e$ in \cite{Ratra96}. Many of the data have 
$\Delta l/l_e< 1/4$ so this is not the sin it may at first
appear. 
\item Compute the weighted mean of the data that fall into each
bin and call that the band-power $\bar{\delta T}$. Use the inverse root of the 
total weight as the error bar. Use all the {\it unique} data in 
\cite{Ratra96}. Be aware that many of the data points are unconfirmed!
\item Compute the arithmetic mean value $\bar l$, of the $l_e$ that fall in a given
bin. 
\item Plot $\bar l$ versus $\bar{\delta T}$ and connect the ends of the 
error bars.
\item Ignore the intrinsic calibration uncertainties and the relative calibration
uncertainties.
\item Ignore upper limits for $l<500$. For some cases,
(eg SASK\cite{SASK}), the data can be combined to give a detection 
\cite{Netterfield97}.
\item In addition to the Ratra compilation, add the  Tegmark
results \cite{Tegmark96}, ATCA \cite{ATCA}, and the new CAT
\cite{CAT} results. 
\end{itemize}

\begin{figure}
\epsfysize=3.5truein
\epsfverbosetrue
\centerline{\epsfbox{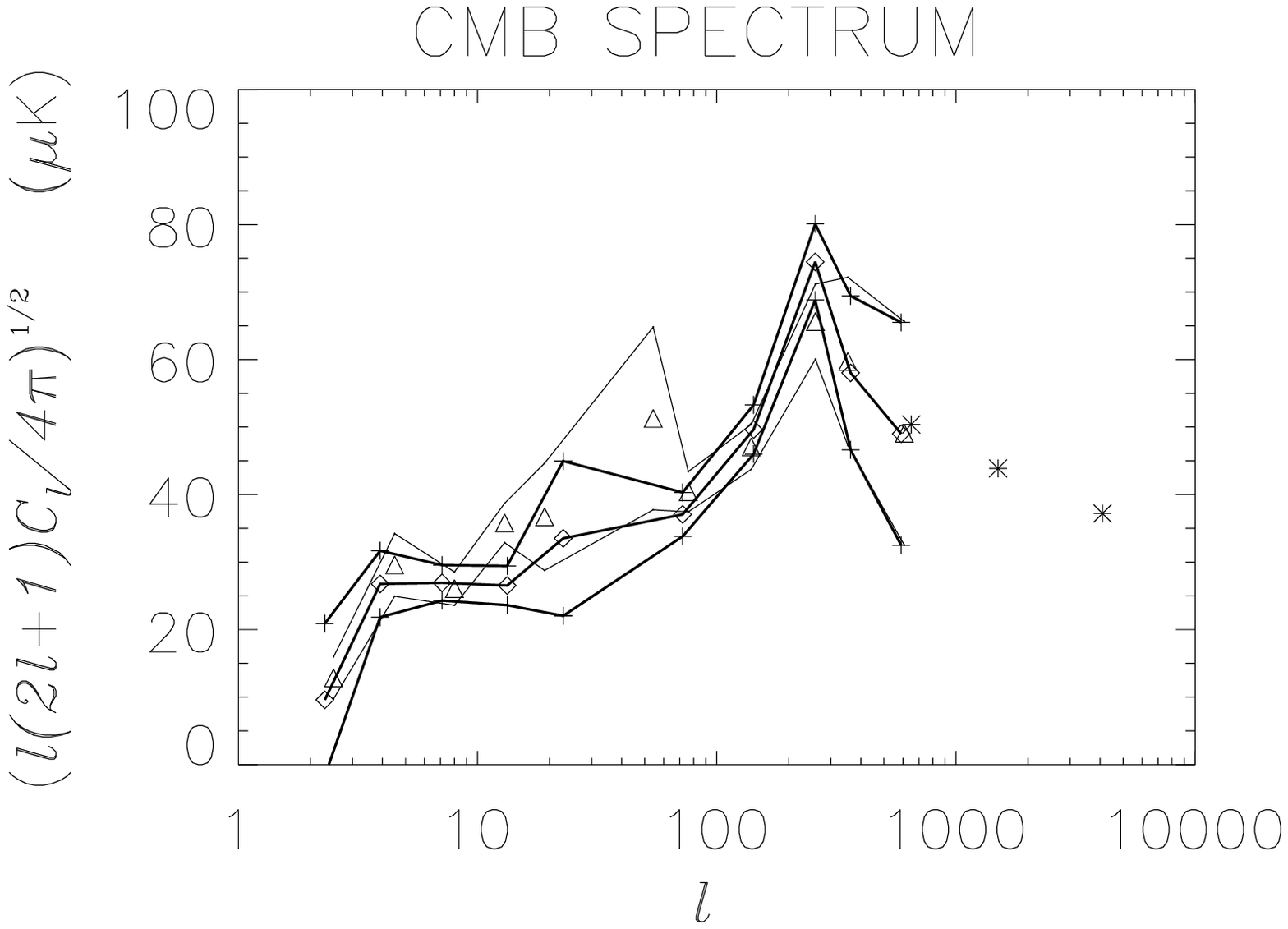}}
\begin{description}
\item[Figure 11.] Angular spectrum of the anisotropy. The thick lines
(diamonds) were obtained following the prescription given in the text.
Bharat Ratra has used somewhat different criteria and obtained
the results indicated by the thin lines. The asterisks near
$l=1000$ are upper limits. Clearly, the data indicate a
rise in the spectrum from $l=10$ to $l\approx 200$.
For $l>20$, the quantity on the y-axis is the same as in Figure
3.  
\end{description}
\end{figure}

The principle conclusion one should draw from Figure 11 is that
there is a general rise in $\delta T_l$ as one moves  from the
{\it COBE} scales to smaller angular scales. This is a stunning
observation that was predicted long before the anisotropy was
discovered. Has a peak to the spectrum been detected? Possibly,
but it is still too early to say this with any confidence.  All
indications are that the power spectrum at $l\approx 10$ is lower
than at $l\approx 200$ but we cannot say where the spectrum turns
over above $l\approx 200$.  The errors are simply too large
and there are too many systematic effects hidden in the data.
Also, there are plenty of examples where 95\% upper limits have
become detections at a higher level and examples where detections
have become upper limits. However, it is somewhat reassuring that the
$\chi^2/(\nu-1)$ for most of the individual bins is not too different from
one.

The measurements have come a long way in the past two years.  In
three cases multiple experiments have observed the same  region
of sky and seen the same thing. They are DMR and FIRS
\cite{Ganga93}, DMR and Tenerife\cite{Tenerife}, and  MSAM
(\cite{Cheng94}, \cite{Cheng96} \& \cite{Inman96}) and SASK
\cite{Netterfield97}, \cite{Page96}. The spectrum of the
fluctuations for MSAM and SASK is thermal from 25 to 200 GHz.
However, one should still view the data with some caution.  In
the analyses that give rise to Figure 11,  an entire data set is
reduced to give one measurement and one statistical error bar.
When this is done,  there is simply not enough signal-to-noise to
quantify systematic effects that are lurking at the  $1\sigma$ to
$2\sigma$ level. On top of this, the analysis of these
experiments is tricky; ``new'' effects are still being discovered
by many groups. Finally, the inter-calibration  of the
experiments is uncertain to the 10\%-15\% level.

To improve on these results, a number of experimental and
observational challenges must be met. The results that went into
Figure 11 are primarily from difference measurements. Eventually
we will want maps of the sky so that experiments are easily
compared, foreground contamination is more easily identified, 
and powerful statistical tests can be performed. Interferometers
offer one proven way to do this at low frequencies (and eventually at
higher frequencies) but other techniques and strategies are
needed. The calibration of the experiments must be better
than 10\% in order to distinguish between the various models.
Currently, the best reference is the intrinsic dipole in the CMB.
Finally, to distinguish features near $l\approx 1000$, high 
resolution will be needed. One desires at least  $\delta
l/l\approx 1/10$. Lower resolutions smear the features in the
power spectrum. 

A measurement of the polarization in the CMB is now within 
grasp. In the standard CDM models, the signal is predicted to be
at 1\% to 5\% of the anisotropy\cite{BE87}. John Ruhl at Santa
Barbara, Suzanne Staggs at Princeton, and Peter Timbie at
Wisconsin are actively working on these measurements. The
polarization is caused by Thompson scattering.  At angular scales
of order ten degrees, the polarization may be used to identify
any primordial gravity waves (tensor modes) though the signal is
expected to  be largest at degree scales. Crittenden and Turok
\cite{Cri95b} point out that there is a correlation between the
polarization and the anisotropy that is different for the scalar
and tensor modes. This correlation should also aid in separating
the CMB polarization from polarized foreground emission, about
which very little is known.

\section{The New Satellite Experiments}

Two satellite missions are planned that will endeavor
to map the CMB anisotropy over the entire sky. The ESA
mission is called {\sl PLANCK} (originally {\sl
COBRAS/SAMBA}). The NASA mission, which is in the
final design and definition phase, is called {\sl MAP}. Because I am
part of the {\sl MAP} team, I will focus on it.

{\sl PLANCK}\footnote{See web site
http://astro.estec.esa.nl/sa-general/projects/cobras/cobras.html
for additional information.} uses both bolometers and HEMT- based
amplifiers. It will carry cryogens so that the bolometers may be
operated at 0.1 K. At this stage in the design, the
instrument is  planned to span the frequencies between roughly 30
and 900 GHz with an angular resolution between $30^{\prime}$ and
$4.4^{\prime}$. The lower frequency channels will be
polarization-sensitive. The planned launch date is 2005.

{\sl MAP} is based on the HEMT amplifiers developed by Marian
Pospieszalski. The radiometers are intrinsically
polarization-sensitive and differential, similar in some regards
to the successful {\sl COBE}/DMR design. The  instrument will
span from 20 to 106 GHz in five frequency bands with an angular
resolution ranging between $54^{\prime}$ and $15^{\prime}$. The
sensitivity per $0.3^{\circ}\times 0.3^{\circ}$ pixel (of which
there are roughly 400,000 in the sky)  will be about $35~\mu$K.
Because the instrument is passively cooled, it can in principle
observe longer than the  15 month design life.

The primary goal of {\sl MAP} is to make multi-frequency,
high-fidelity, high-sensitivity maps of the sky. This requires
extreme control of systematic effects. We believe the best
vantage for these observations is L2, the Earth-Sun Lagrange
point. At L2, the Sun, Earth, and moon are $\approx 90^{\circ}$
out of the beams and the environment is essentially isothermal.
From the work on DMR and balloons, the team has found that 
successful map production requires reference of one pixel to
another over many directions and over many time scales. {\sl MAP}
plans to do this with the scan strategy shown in Figure 12.

\begin{figure}
\epsfysize=4.5truein
\epsfverbosetrue
\centerline{\epsfbox{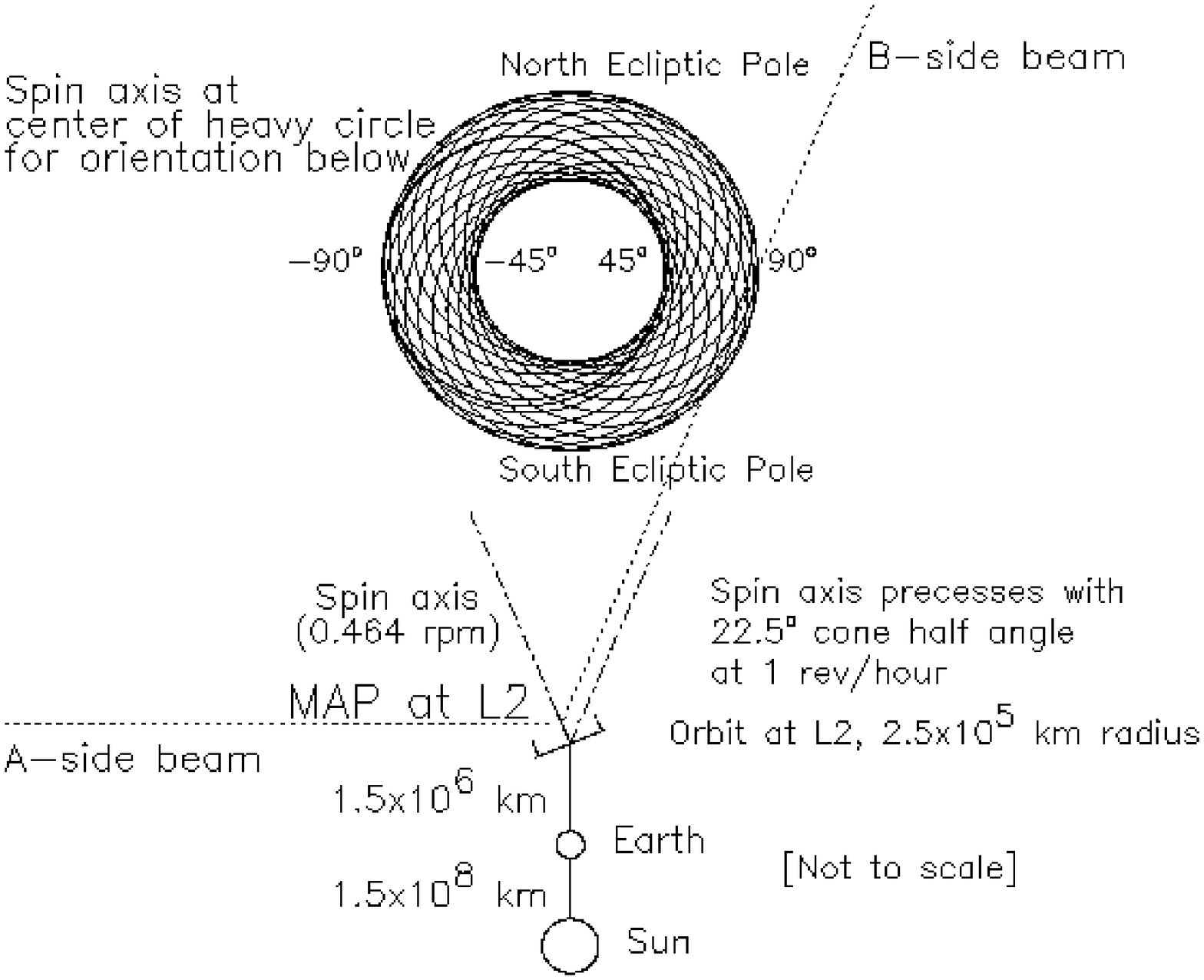}}
\begin{description}
\item[Figure 12.] The {\sl MAP} scan pattern for one hour of observation.
The lines show the path for one side of a differential pair. The
other pair member follows a similar path, only delayed by 1.1
min. There are four principal time scales for the observations.
The phase of the difference signal is switched by $180^{\circ}$
at 2.5 KHz. The spacecraft spins around its symmetry axis with a
2.2 min period (bold circle) with cone opening angle of
roughly $135^{\circ}$. This pattern precesses about the Earth-Sun
line with a period of 60 minutes. Thus, in about 1 hour, over 30\% of
the sky is covered. Every six months, the whole sky is observed.
Note that any pixel is differenced to another pixel in many
directions.
\end{description}
\end{figure}

From a high-quality map, one may not only obtain the power
spectrum, but may also compare the data to the results of
other CMB experiments and to maps of the foreground emission at
different frequencies. Also, a map gives the best data set for 
testing the underlying statistics of the fluctuations. For
instance, we will be able to tell from the {\sl MAP} data if the CMB is
a Gaussian random field.  Finally, with its high sensitivity and
large scale coverage, the time-line data from {\sl MAP} will be
ideal for  searching for transient radio emission.  

The question of how well one can determine the parameters of
cosmological models is still an active area of research. The most
recent published work on parameter estimation for inflation-based
models is in \cite{Jungman96} but one must remember that there
are other classes of promising models. At this school, Dick
Bond discussed an approach where one works in an eigen-parameter
space to circumvent the strong covariance between many of the 
standard parameters such as $\Omega_B$, $\Lambda$, $h$, etc. At
any rate, if the anisotropy is normally distributed, the {\sl
MAP} CMB data will be cosmic variance limited up to $l\approx
600$  (assuming the foreground/radio source emission is
successfully removed) and will probe multipoles up to $l\approx
1000$. 

In the current schedule, the satellite design and definition will
be complete by November 1997 and then the building will begin.
{\sl MAP} is scheduled for launch late in 2000. The {\sl MAP} science
team is comprised of Chuck Bennett (PI) at NASA/GSFC, Mark
Halpern at UBC, Gary Hinshaw at NASA/GSFC, Norm Jarosik at
Princeton, John Mather at NASA/GSFC, Steve Meyer at Chicago,
Lyman Page at Princeton, Dave Spergel at Princeton, Dave
Wilkinson at Princeton, and Ned Wright at UCLA. More information
about {\sl MAP}, the CMB, and other experiments may be obtained
from http://map.gsfc.nasa.gov/.

I would like to thank Roberta Bernstein, Venya Berezinsky, Piero
Gallotti, David Schramm and the Ettore Majorana Center staff for
organizing a wonderful school. Marsala will never taste the same.
Conversations with many colleagues were helpful in preparing
these notes. I would especially like to thank Tom Herbig, Gary
Hinshaw, Bharat Ratra, Suzanne Staggs, and Ned Wright. Ned gave me the computer
code to produce Figure 4. This work was supported by the US
National Science Foundation and the  David and Lucile Packard
Foundation.



\begin{thebibliography}{}
\bibitem{ACE} ACE and BEAST. These two new projects are aimed at
using HEMTs between 26 and 100 GHz on both super-pressure
and conventional long-duration balloon platforms. The finest
angular resolution will be near $1/5^{\circ}$.
\bibitem{BOOMERanG}BOOMERanG is a collaboration between the Caltech,
Berkeley, Santa Barbara (Ruhl) and Rome groups.  It will use
bolometers to measure the anisotropy in the CMB between 90 and
410 GHz. The ultimate goal is a circumpolar Antarctic flight.
\bibitem{CBI}CBI: Cosmic Background Imager. This is an interferometer
that plans to produce maps of the microwave sky near 30 GHz.
\bibitem{MAT}MAT. This is the Mobile Anisotropy Telescope. It is
similar to QMAP but is designed to operate from the ground
in Chile.
\bibitem{OVRO}OVRO. The Owen's Valley Radio Observatoty telescopes 
operate with various receivers between
15 and 30 GHz. The 40 m dish has a $2^{\prime}$ beam, and the 
5.5 m has a $7.3^{\prime}$ beam. The experiments are aimed
primarily at small angular scales.
\bibitem{QMAP}QMAP. This is a balloon-borne experiment that uses a
combination of HEMTs and SIS detectors. The angular resolution is
$1/5^{\circ}$. This experiment is designed to produce
``true'' maps of the sky.
\bibitem{SASK}SASK. These experiments are based on HEMT amplifiers 
operating between 26 and 46 GHz. They were performed in
Saskatoon, Saskatchewan CA. Three years of observations have gone into 
the final data set.
\bibitem{SuZie}SuZIE is a bolometer-based experiment that
observes from the ground. It is primarily intended to measure the
SZ effect at high frequencies though it will also give information
on the anisotropy at small scales.
\bibitem{VCA}VCA: Very Compact Array. This is an interferometer
being developed at the University of Chicago. It will produce
maps of the CMB at 30 GHz and be sensitive to larger angular
scales than CBI.
\end{thebibliography}
\end{document}